\documentstyle[12pt]{article}

\begin{document}

\begin{flushright}
DPUR/TH/68\\
May, 2020\\
\end{flushright}
\vspace{20pt}

\pagestyle{empty}
\baselineskip15pt

\begin{center}
{\large\bf  On Restricted Weyl Symmetry
\vskip 1mm }

\vspace{20mm}

Ichiro Oda\footnote{
           E-mail address:\ ioda@sci.u-ryukyu.ac.jp
                  }

\vspace{10mm}
           Department of Physics, Faculty of Science, University of the 
           Ryukyus,\\
           Nishihara, Okinawa 903-0213, Japan\\

\end{center}


\vspace{10mm}
\begin{abstract}

We discuss the physics of a restricted Weyl symmetry in a curved space-time 
where a gauge parameter $\Omega(x)$ of Weyl transformation satisfies a constraint 
$\Box \Omega = 0$. First, we present a model of QED where we have a restricted gauge 
symmetry in the sense that a $U(1)$ gauge parameter $\theta(x)$ obeys a similar 
constraint $\Box \theta = 0$ in a flat Minkowski space-time. Next, it is precisely shown 
that a global scale symmetry must be spontaneously broken at the quantum level. 
Finally, we discuss the origin of the restricted Weyl symmetry and show that its symmetry 
can be derived from a full Weyl symmetry by taking a gauge condition $R = 0$ in the BRST 
formalism.     

\end{abstract}

\newpage
\pagestyle{plain}
\pagenumbering{arabic}


\section{Introduction}

Both local and global scale symmetries are very mysterious symmetries in that they
are ubiquitous in nature from particle physics to cosmology, but in the real world 
they usually emerge as approximate symmetries which are broken either explicitly 
by anomalies or badly by the presence of a built-in scale in theories.\footnote{We 
sometimes call a local and global scale symmetry a Weyl symmetry and scale symmetry, 
respectively \cite{Weyl, O'Raifeartaigh, Fujii}.} If the scale 
symmetries are exact ones which are only spontaneously broken as in the gauge symmetries 
in the standard model (SM) of particle physics, they might shed some light on various important 
unsolved problems such as cosmological constant problem and the gauge hierarchy problem etc. 

In this short article, we explore an idea that there might be an intermediate scale symmetry  
between local and global scale symmetries, which is dubbed a restricted Weyl symmetry in a curved space-time 
\cite{Edery1, Edery2, Edery3}. In the restricted Weyl symmetry, a gauge parameter, which is nothing 
but a conformal factor $\Omega(x)$, is constrained by a condition $\Box \Omega = 0$ whereas in a conventional or 
full Weyl transformation the conformal factor is an unconstrained and free parameter.   
In the restricted Weyl symmetry, we are allowed to work with a generic dimensionless action
as in a scale symmetry, which should be contrasted to the situation in the full Weyl symmetry where only the 
conformal tensor squared is an invariant action if we neglect the other fields except for the metric $g_{\mu\nu}$.

The structure of this article is the following: In Section 2, we review a restricted gauge symmetry in QED which appears 
after we fix the gauge invariance by the Lorenz condition, i.e., the Lorenz gauge.  The restricted gauge symmetry resembles
the restricted Weyl symmetry in the sense that both the symmetries are constrained by an equation $\Box \Theta = 0$ 
where $\Box$ denotes the d'Alembertian operator and $\Theta$ stands for a generic gauge parameter.
Of course, an obvious difference between the restricted gauge symmetry and the restricted Weyl one lies in the fact 
that the former is defined in a flat Minkowski space-time while the latter is so in a curved Riemannian space-time.
Their similarity, however, gives us some hints about the spontaneous symmetry breakdown and the origin of the restricted
Weyl symmetry, which are dealt with in Section 3 and 4, respectively. The final section is devoted to conclusion.

\section{Restricted gauge symmetry}

Let us start with QED which is gauge-fixed by the Lorenz gauge:
\begin{eqnarray}
{\cal{L}} = - \frac{1}{4} F_{\mu\nu}^2 + \bar \psi ( i \gamma^\mu \partial_\mu - m ) \psi
+ e A_\mu \bar \psi \gamma^\mu \psi + B \partial_\mu A^\mu + \frac{\alpha}{2} B^2,
\label{QED}  
\end{eqnarray}
where $A_\mu, \psi$ and $B$ are respectively the electromagnetic field, spinor field and Nakanishi-Lautrup field,
the field strength is defined as $F_{\mu\nu} \equiv \partial_\mu A_\nu - \partial_\nu A_\mu$, and $\alpha$ is a real number
which can be chosen for our convenience. This Lagrangian density has a restricted $U(1)$ gauge invariance given by
\begin{eqnarray}
\delta A_\mu = \partial_\mu \theta, \quad 
\delta \psi = i e \theta \psi, \quad
\delta \bar \psi = - i e \theta \bar \psi, \quad 
\delta B = 0,
\label{Res-Gauge}  
\end{eqnarray}
where $\Box \theta = 0$ where $\Box = \eta^{\mu\nu} \partial_\mu \partial_\nu$ is the d'Alembertian operator
in a flat Minkowski space-time.

Then, let us ask ourselves what solution to the constraint $\Box \theta (x) = 0$ is.
Since $\Box \theta = 0$ is nothing but the Klein-Gordon equation for a massless real scalar 
field, a general solution is given by
\begin{eqnarray}
\theta (x) = \int \frac{d^3 k}{\sqrt{(2 \pi)^3 2 k_0}} [ a(k) e^{-ikx}
+ a^\dagger(k) e^{ikx} ],
\label{Omega}  
\end{eqnarray}
where $k_0 = |\vec{k}|$. In the absence of the spinor field $\psi = 0$ and with the $\alpha = 1$
gauge, the gauge field also obeys the same field equation $\Box A_\mu = 0$ as in $\Box \theta = 0$, 
so we can then gauge away one component in $A_\mu$ by means of the residual symmetry (\ref{Res-Gauge}).
Incidentally, if we are interested in only respecting the restricted gauge invariance, and we do not inquire 
its origin and ignore an issue of non-renormalizability, we could add any invariant terms such as $( \partial_\mu A^\mu )^n$ 
with integers $n \ge 2$ to the Lagrangian density (\ref{QED}).

Now we are interested in only zero-mode solutions.
The constraint $\Box \theta = 0$ is then easily solved to be \cite{Kugo}
\begin{eqnarray}
\theta = a_\mu x^\mu + b,  
\label{Theta-constraint}  
\end{eqnarray}
where $a_\mu, b$ are infinitesimal constants. With this solution, the restricted gauge transformation 
can be rewritten as
\begin{eqnarray}
\delta A_\mu = a_\mu, \quad 
\delta \psi = i e (a_\mu x^\mu + b) \psi, \quad
\delta \bar \psi = - i e (a_\mu x^\mu + b) \bar \psi, \quad
\delta B = 0.
\label{Global-Res-Gauge}  
\end{eqnarray}
Since the transformation (\ref{Global-Res-Gauge}) is a global symmetry, following the Noether
theorem \cite{Noether} we can derive the Noether currents $j_\mu \,^\rho, j^\rho$ corresponding to 
the parameters $a_\mu, b$, respectively:
\begin{eqnarray}
j_\mu \,^\rho = F_\mu \,^\rho + \delta_\mu^\rho B - x_\mu e \bar \psi \gamma^\rho \psi, \qquad 
j^\rho = - e \bar \psi \gamma^\rho \psi. 
\label{Noether1}  
\end{eqnarray}
The Noether charges $Q_\mu, Q$ are respectively defined as
\begin{eqnarray}
Q_\mu = \int d^3 x \ j_\mu \,^0, \qquad 
Q = \int d^3 x \ j^0. 
\label{Noether-charge1}  
\end{eqnarray}

Next, let us show that a vectorial charge $Q_\mu$ is necessarily broken spontaneously \cite{Kugo}. 
Actually, we find that  
\begin{eqnarray}
\delta A_\nu = [ i ( a^\mu Q_\mu + b Q ), A_\nu ] = \partial_\nu \theta = a_\nu,
\label{Inf-transf}  
\end{eqnarray}
which implies that 
\begin{eqnarray}
[ i Q_\mu, A_\nu ] = \eta_{\mu\nu}, \qquad 
[ i Q, A_\nu ] = 0. 
\label{CR1}  
\end{eqnarray}
Taking the vacuum expectation value of the former relation leads to
\begin{eqnarray}
\langle 0 | [ i Q_\mu, A_\nu ] | 0 \rangle = \eta_{\mu\nu}. 
\label{VEV-CR1}  
\end{eqnarray}
Eq. (\ref{VEV-CR1}) means that a global symmetry generated by $Q_\mu$ is
spontaneously broken. As a result, $A_\mu$ includes a massless Nambu-Goldstone (NG)
boson, but it may be either a scalar or a vector particle. It has been already proved that
each possibility exactly corresponds to either a symmetry generated by $Q$ is
spontaneously broken or the symmetry remains unbroken \cite{Kugo}. In particular, 
in the latter case we can regard the photon as a NG boson coming from the SSB of 
$Q_\mu$ \cite{Ferrari}.

\section{Restricted Weyl symmetry}

In this section, we consider a gravitational theory coupled to a $U(1)$ gauge theory with a complex 
scalar field where there is a restricted Weyl invariance in the sense that a gauge parameter of 
Weyl transformation, $\Omega(x)$, satisfies a constraint, $\Box \Omega(x) = 0 $\footnote{We follow 
the conventions and notation of the MTW textbook \cite{MTW}.}:
\begin{eqnarray}
{\cal{L}} = \sqrt{-g} \left( \alpha R^2 - \xi R |\Phi|^2 - |D_\mu \Phi|^2 - \lambda |\Phi|^4 
- \frac{1}{4} F_{\mu\nu} F^{\mu\nu} \right),
\label{Weyl Lag0}  
\end{eqnarray}
where the covariant derivative is defined as $D_\mu \Phi \equiv ( \partial_\mu - i e A_\mu ) \Phi$.
Without loss of generality, in this article we drop the gauge field $A_\mu$ and we work with the
following Lagrangian density \cite{Edery2}:
\begin{eqnarray}
{\cal{L}} = \sqrt{-g} \left( \alpha R^2 - \xi R |\Phi|^2 - |\partial_\mu \Phi|^2 
- \lambda |\Phi|^4 \right).
\label{Weyl Lag}  
\end{eqnarray}
Indeed, this Lagrangian density is invariant under the restricted Weyl transformation  
\begin{eqnarray}
g_{\mu\nu} \rightarrow g^\prime_{\mu\nu} = \Omega^2 (x) g_{\mu\nu}, \qquad 
\Phi \rightarrow \Phi^\prime = \Omega^{-1}(x) \Phi, 
\label{Res-Weyl}  
\end{eqnarray}
where the gauge parameter obeys a constraint $\Box \Omega = 0$. In order to prove the
invariance, we need to use the following transformation of the scalar curvature 
under (\ref{Res-Weyl}):
\begin{eqnarray}
R \rightarrow R^\prime = \Omega^{-2} ( R - 6 \Omega^{-1} \Box \Omega ). 
\label{Weyl-R}  
\end{eqnarray}

We are at present interested in zero-mode solution to the equation $\Box \Omega = 0$.
It is obvious that $\Omega(x) = const.$ is a zero mode solution corresponding to a global scale 
invariance. Then, let us pay our attention to the scale invariance. In an infinitesimal form $\Omega 
= e^\Lambda$ with $|\Lambda| \ll 1$, the infinitesimal gauge parameter $\Lambda$ must obey 
a constraint $\Box \Lambda = 0$ as well, so the zero-mode solution is given by 
\begin{eqnarray}
\Lambda (x) = b,
\label{Lambda}  
\end{eqnarray}
where $b$ is a constant. 

Since there is a global invariance associated with the parameter $b$, we can construct a conserved 
Noether charge $Q$. To derive a conserved current according to the Noether theorem \cite{Noether}, 
let us first recall that the Lagrangian density is assumed to contain only up to the first derivatives of fields.
For this aim, we will first rewrite a $\alpha R^2$ term into the form $\varphi R - \frac{1}{4 \alpha} \varphi^2$
where $\varphi$ is a scalar field with the dimension of mass squared:
\begin{eqnarray}
{\cal{L}} = \sqrt{-g} \left( \varphi R - \frac{1}{4 \alpha} \varphi^2 - \xi R |\Phi|^2 - |\partial_\mu \Phi|^2 
- \lambda |\Phi|^4 \right).
\label{Weyl Lag2}  
\end{eqnarray} 
Then, we will perform the integration by parts to make the second derivative in $R$ change the first derivative.
Following the calculation \cite{Fujii, Oda-H}, it turns out that the conserved current for dilatation reads
\begin{eqnarray}
J^\mu = \sqrt{-g} g^{\mu\nu} \partial_\nu [ 6 \varphi + ( 1 - 6 \xi ) |\Phi|^2 ].
\label{Noether current}  
\end{eqnarray} 
We can also verify that this current is conserved, $\partial_\mu J^\mu = 0$, by using the field equations
as desired. One might wonder why no derivatives of the metric appear in the expression of $J^\mu$.
This is because the derivatives of $\varphi$ and $\Phi$ are mixed with the metric, thus making $J_0$
serve as a generator of the metric transformation. Moreover, in case of a conformal coupling $\xi 
= \frac{1}{6}$ and the absence of a $R^2$ term, the conserved current is identically vanishing 
\cite{Jackiw, Oda-U}. In other words, the nonvanishing current requires us to treat with a $R^2$ term and/or 
a scalar matter field $\Phi$ with a non-conformal coupling $\xi \neq \frac{1}{6}$. 
 
Using the corresponding Noether charge defined as $Q = \int d^3 x J^0$, we find that 
\begin{eqnarray}
\delta g_{\mu\nu}  = [ i b Q, g_{\mu\nu} ] = 2 b g_{\mu\nu},
\label{Q-transf1}  
\end{eqnarray}
from which we have
\begin{eqnarray}
[ i Q, g_{\mu\nu} ] = 2 g_{\mu\nu},
\label{Q-CR1}  
\end{eqnarray}
Assuming $\langle 0| g_{\mu\nu} |0 \rangle = \eta_{\mu\nu}$\footnote{Of course, we can consider a more 
general fixed background $\bar g_{\mu\nu}$ which satisfies $\langle 0| g_{\mu\nu} |0 \rangle = \bar g_{\mu\nu}$, 
but a flat Minkowski background assures that a $GL(4)$ symmetry is spontaneously broken to an $SO(1, 3)$ 
and consequently the graviton is a Nambu-Goldstone tensor boson in quantum gravity \cite{Nakanishi}.} and taking 
the vacuum expectation value of Eq. (\ref{Q-CR1}) leads to
\begin{eqnarray}
\langle 0| [ i Q, g_{\mu\nu} ] |0 \rangle = 2 \eta_{\mu\nu}.
\label{Q-SSB}  
\end{eqnarray}
Eq. (\ref{Q-SSB}) clearly implies 
that the global scale invariance must be broken spontaneously at the quantum level \cite{Oda-S}.  
Note that this finding is obtained by using the zero-mode solution to the constraint $\Box \Omega = 0$.

Next, let us verify explicitly that this is the case by moving from the Jordan frame to the Einstein frame
\cite{Oda-C}. To do so, we will move to the Einstein frame by implementing a local conformal transformation 
\begin{eqnarray}
g_{\mu\nu} \rightarrow g_{\ast\mu\nu} = \Omega^2 (x) g_{\mu\nu}, \qquad 
\Phi \rightarrow \Phi_\ast = \Omega^{-1}(x) \Phi.
\label{Transf-E-frame}  
\end{eqnarray}
Under this conformal transformation we have \cite{Fujii}
\begin{eqnarray}
\sqrt{-g} = \Omega^{-4} \sqrt{-g_\ast}, \qquad
R = \Omega^2 ( R_\ast + 6 \Box_\ast f - 6 g_\ast^{\mu\nu} f_\mu f_\nu ), 
\label{Transf-E-frame2}  
\end{eqnarray}
where we have defined
\begin{eqnarray}
f \equiv \log \Omega, \quad
\Box_\ast f \equiv \frac{1}{\sqrt{- g_\ast}} \partial_\mu ( \sqrt{- g_\ast} g_\ast^{\mu\nu} \partial_\nu f), \quad
f_\mu \equiv \partial_\mu f = \frac{\partial_\mu \Omega}{\Omega}.
\label{f}  
\end{eqnarray}

Then, the Lagrangian density (\ref{Weyl Lag2}) is cast to the form
\begin{eqnarray}
{\cal{L}} &=& \sqrt{-g_\ast} \Biggl[ ( \varphi \Omega^{-2} - \xi |\Phi_\ast|^2 ) 
( R_\ast + 6 \Box_\ast f - 6 g_\ast^{\mu\nu} f_\mu f_\nu ) - \frac{1}{4 \alpha} \varphi^2 \Omega^{-4}  
\nonumber\\
&-& \Omega^{-2} g_\ast^{\mu\nu} \partial_\mu (\Omega \Phi_\ast^\dagger) \partial_\nu 
(\Omega \Phi_\ast) - \lambda |\Phi_\ast|^4  \Biggr].
\label{E-Lag}  
\end{eqnarray}
To reach the Einstein frame, we have to choose a conformal factor $\Omega(x)$ to satisfy 
a relation
\begin{eqnarray}
\varphi \Omega^{-2} = \xi |\Phi_\ast|^2 + \frac{M_{Pl}^2}{2},
\label{Planck-mass}  
\end{eqnarray}
where $M_{Pl}$ is the reduced Planck mass. As a result, with a redefinition $\omega(x) \equiv \sqrt{6}
M_{Pl} f(x)$, we obtain a Lagrangian density in the Einstein frame:
\begin{eqnarray}
{\cal{L}} &=& \sqrt{-g_\ast} \Biggl[ \frac{M_{Pl}^2}{2} R_\ast - \frac{1}{2} g_\ast^{\mu\nu} 
\partial_\mu \omega \partial_\nu \omega - \frac{1}{16 \alpha} M_{Pl}^4
- |\partial_\mu \Phi_\ast|^2 - \frac{\xi}{4 \alpha} M_{Pl}^2 |\Phi_\ast|^2
\nonumber\\
&-& \left( \lambda + \frac{\xi^2}{4 \alpha} \right) |\Phi_\ast|^4 
+ \left( \frac{1}{\sqrt{6} M_{Pl}} \Box_\ast \omega - \frac{1}{6 M_{Pl}^2} g_\ast^{\mu\nu} 
\partial_\mu \omega \partial_\nu \omega \right) |\Phi_\ast|^2 \Biggr].
\label{E-Lag2}  
\end{eqnarray}
It is worthwhile to notice that spontaneous symmetry breakdown for a scale invariance
has occurred and consequently we have a massless Nambu-Goldstone boson $\omega(x)$, which is
often called ``dilaton''. As a bonus, a gauge symmetry is also spontaneously broken if we choose 
the parameters to be 
\begin{eqnarray}
\frac{\xi}{4 \alpha} < 0, \qquad \lambda + \frac{\xi}{4 \alpha} > 0.
\label{Parameter}  
\end{eqnarray}
Also note that the last two nonrenormalizable terms in Eq. (\ref{E-Lag2}) are suppressed by the
Planck mass so that they would make only a small contribution at low energies $E \ll M_{Pl}$.

\section{Origin of Restricted Weyl symmetry}

We wish to understand the origin of a restricted Weyl symmetry. We will see that the restricted Weyl symmetry
emerges as a residual symmetry of Weyl symmetry in a similar way that a restricted gauge symmetry appears
as a residual symmetry of the conventional gauge symmetry after we take a Lorenz gauge in QED. 

For generality, let us work with a general theory which is invariant under a full Weyl transformation:
\begin{eqnarray}
{\cal{L}} = \sqrt{-g} \left( \frac{1}{2} \partial_\mu \phi \partial^\mu \phi + \frac{1}{12} R \phi^2
+ c_1 C_{\mu\nu\rho\sigma} C^{\mu\nu\rho\sigma} - \frac{\lambda}{4 !} \phi^4 \right),
\label{General-Lag}  
\end{eqnarray}
where $\phi$ denotes a (ghost-like) scalar field, $C_{\mu\nu\rho\sigma}$ is a conformal tensor
and $c_1$ is a constant. It is well-known that this Lagrangian density is invariant under a Weyl transformation 
without an additional constraint on $\Omega(x)$:
\begin{eqnarray}
g_{\mu\nu} \rightarrow g^\prime_{\mu\nu} = \Omega^2 (x) g_{\mu\nu}, \qquad 
\phi \rightarrow \phi^\prime = \Omega^{-1}(x) \phi. 
\label{Full-Weyl}  
\end{eqnarray}

In order to derive the restricted Weyl invariance from the full Weyl invariance, one has to fix the
Weyl symmetry in such a way that the gauge fixing condition breaks the full Weyl invariance but
leaves the restricted Weyl invariance unbroken. A suitable gauge condition is $R = 0$.\footnote{We
could take a more general gauge condition $R + k \phi^2 = 0$ ($k$ is a constant) if necessary
\cite{Oda-S, Oda-E}.} This gauge choice can be achieved as follows: Let us expand $g_{\mu\nu}$
around a flat metric $\eta_{\mu\nu}$ as $g_{\mu\nu} = \eta_{\mu\nu} + h_{\mu\nu}$ with $h_{\mu\nu}$ being
a small fluctuation ($|h_{\mu\nu}| \ll 0$). In order to show that we can take a gauge condition $R = 0$ 
for the full Weyl symmetry, we start with $R \neq 0$, and then show that we can arrive at $R^\prime = 0$
by means of a Weyl transformation as seen in the relation (\ref{Weyl-R}). With the expansion $g_{\mu\nu} 
= \eta_{\mu\nu} + h_{\mu\nu}$, the scalar curvature reads $R = - \Box h$ ($h \equiv \eta^{\mu\nu} 
h_{\mu\nu}$) to the linear order in $h_{\mu\nu}$.  Using an infinitesimal Weyl transformation 
$\Omega = e^\Lambda$, the RHS of Eq. (\ref{Weyl-R}) can be rewritten to the linear order in $h_{\mu\nu}$
and $\Lambda$ as 
\begin{eqnarray}
\Omega^{-2} ( R - 6 \Omega^{-1} \Box \Omega ) = - \Box ( h + 6 \Lambda ).
\label{R-gauge}  
\end{eqnarray}
Thus, if we choose the infinitesimal gauge parameter as $\Lambda = - \frac{1}{6} h$, we can certainly
achieve $R^\prime = 0$.

We therefore attempt to fix the Weyl invariance in terms of a gauge condition $R = 0$ in the BRST formalism. 
First of all, the BRST transformation for the Weyl symmetry reads
\begin{eqnarray}
\delta_B g_{\mu\nu} &=& 2 c g_{\mu\nu}, \qquad
\delta_B \sqrt{-g} = 4 c \sqrt{-g}, \qquad
\delta_B R = - 2 c R - 6 \Box c,
\nonumber\\
\delta \phi &=& - c \phi, \qquad \delta_B \bar c = i B, \qquad \delta_B c = \delta_B B = 0.
\label{BRST}  
\end{eqnarray}

Next, a Lagrangian density for the gauge condition and the FP ghost is of form
\begin{eqnarray}
{\cal{L}}_{GF+FP} &=& - i \delta_B \left[ \sqrt{-g} \bar c \left( R + \frac{\alpha}{2} B \right) \right]
\nonumber\\
&=& \sqrt{-g} \left( \hat B R + \frac{\alpha}{2} \hat B^2 - 6 i \bar c \Box c \right)
\nonumber\\
&=& \sqrt{-g} \left( - \frac{1}{2 \alpha} R^2 + 6 i g^{\mu\nu} \partial_\mu \bar c \partial_\nu c \right),
\label{GF-FP}  
\end{eqnarray}
where we have defined as $\hat B \equiv B + 2 i \bar c c$ and in the last step we performed the
path integral over the auxiliary field $\hat B$ and integration by parts \cite{Oda-E}.

Thus, we arrive at a gauge-fixed and BRST-invariant Lagrangian density given by
\begin{eqnarray}
{\cal{L}} &=& \sqrt{-g} \Biggl( \frac{1}{2} \partial_\mu \phi \partial^\mu \phi + \frac{1}{12} R \phi^2
+ c_1 C_{\mu\nu\rho\sigma} C^{\mu\nu\rho\sigma} - \frac{\lambda}{4 !} \phi^4 
\nonumber\\
&-& \frac{1}{2 \alpha} R^2 + 6 i g^{\mu\nu} \partial_\mu \bar c \partial_\nu c \Biggr).
\label{BRST-Lag}  
\end{eqnarray}
It is worthwhile to note that this BRST-invariant Lagrangian density is also invariant under the
restricted Weyl transformation. Actually, the first three terms are manifestly invariant under the
restricted Weyl transformation since they are so under the full Weyl transformation. The last
two terms turn out to be invariant under not the full Weyl transformation but the restricted Weyl 
transformation. For instance, the invariance of the ghost term can be shown as
follows: First,  let us assume that both FP-ghost and FP-antighost have the Weyl weight $-1$, that is,
under the Weyl transformation they transform as
\begin{eqnarray}
c \rightarrow c^\prime = \Omega^{-1} (x) c, \qquad 
\bar c \rightarrow \bar c^\prime = \Omega^{-1} (x) \bar c. 
\label{Full-Weyl2}  
\end{eqnarray}
Then, we find that under the Weyl transformation the ghost term transforms as
\begin{eqnarray}
&{}& \sqrt{-g} g^{\mu\nu} \partial_\mu \bar c \partial_\nu c 
\nonumber\\
&\rightarrow& \sqrt{-g} g^{\mu\nu} \left[ \partial_\mu \bar c \partial_\nu c
+ \Omega^{-1} \nabla_\mu \nabla_\nu \Omega \cdot \bar c c
- \nabla_\mu (\Omega^{-1} \nabla_\nu \Omega \cdot \bar c c) \right]. 
\label{Ghost-transf}  
\end{eqnarray}
Hence, the ghost kinetic term is invariant under the restricted Weyl transformation up to a surface
term.
In this way, we have succeeded in deriving the restricted Weyl invariance by beginning with
a full Weyl invariance by taking a gauge condition $R = 0$. Of course, if we do not pay an
attention to the origin of the restricted Weyl invariance, we can add any terms which are invariant under 
the restricted Weyl symmetry to (\ref{BRST-Lag}), and remove the ghost term from (\ref{BRST-Lag}). 
However, the fact that the restricted
Weyl symmetry can be derived from the full Weyl symmetry not only sheds some light on its geometrical
structure but also clarifies that the restricted Weyl symmetry is not an ${\it{ad \ hoc}}$ but natural symmetry.

\section{Conclusion}

In this article, we have investigated a restricted Weyl symmetry. In particular, we have clarified two points:
First, on the basis of the BRST formalism we have shown that a global scale invariance, which is included
in the restricted Weyl invariance, must be broken spontaneously at the quantum level. Second, we have 
derived a gauge-fixed and BRST-invariant action, which is also invariant under the restricted Weyl symmetry
as well as diffeomorphisms, by starting with a Weyl-invariant gravitational theory by fixing the Weyl symmetry 
by a gauge condition $R = 0$. Our derivation clarifies the origin of the restricted Weyl symmetry. 

Nevertheless, there seem to remain many unsolved problems relevant to the restricted Weyl symmetry. 
In what follows, we will comment on only two important problems to be understood in future.   
One of them is related to trace anomaly. It has been recently established that a scale symmetry is a quantum
symmetry which is broken only spontaneously and is free from trace anomaly \cite{Englert, Shaposhnikov,
Ghilencea}. This is done by using a subtraction function $\mu(\omega)$ ($\omega$ is a dilaton in Section 3)
instead of a dimensionful subtraction scale $\mu$ in the dimensional regularization method. 

However, as a price we have to pay, an infinite number of counter-terms like $\left( \frac{|\Phi|}{\omega} \right)^n$
with $n$ being integers would be needed, thereby breaking the property of renormalizability. This issue of
nonrenormalizability is not so serious in the context at hand since the Einstein-Hilbert term in  (\ref{E-Lag2})
is a nonrenormalizable term as well. The model made in Section 3 could be a candidate of physics beyond 
the standard model (BSM) so one should construct such a BSM in an explicit manner in terms of the manifestly 
scale invariant regularization scheme to attack the gauge hierarchy problem and the cosmological constant problem etc.

The other problem is related to a constraint $\Box \Omega = 0$ in a restricted Weyl symmetry.
As reviewed in Section 1, a constraint $\Box \theta = 0$ in QED removes one dynamical degree of freedom in $A_\mu$.
It is unclear what dynamical degree of freedom can be removed by the constraint $\Box \Omega = 0$ in the case 
of the restricted Weyl symmetry. It is known that the constraint $\Box \Omega = 0$ has an infinite number of 
classical solutions, so it should remove one dynamical degree of freedom. To put differently, we do not understand 
the physical meaning of the constraint $\Box \Omega = 0$ yet. We wish to return these problems in future.

\begin{flushleft}
{\bf Acknowledgements}
\end{flushleft}

We would like to thank T. Kugo for valuable discussions and reading of this manuscript.


\end{document}